\documentstyle[11pt,paspconf,epsf]{article}
\def\arg#1{{\it#1\/}}

\def\edcomment#1{\iffalse\marginpar{\raggedright\sl#1\/}\else\relax\fi}
\reversemarginpar

\begin{document}
\title{Systematic Variations in Age and Metallicity Along the
	Early-Type Galaxy Sequence}
\author{Michael A. Pahre}
\affil{Harvard-Smithsonian Center for Astrophysics}
\author{S. G. Djorgovski}
\affil{Palomar Observatory, California Institute of Technology}
\author{Reinaldo R. de Carvalho}
\affil{Departamento de Astrofisica, Observatorio Nacional, CNPq, Brazil}

\begin{abstract}
The form of the early-type galaxy scaling relation (the Fundamental Plane 
or FP) is a direct indicator of the underlying physical origins for the 
galaxy sequence.
Observed properties of the FP include:
(1) the slope increases with wavelength;
(2) the slope deviates from the virial expectation (assuming homology and 
	constant $M/L$) at all wavelengths;
(3) the intercept evolves passively with redshift; and
(4) the slope decreases slowly with redshift.
The first property implies that stellar populations contribute to the slope of
the FP, the second and fourth properties exclude metallicity effects as the sole
cause of the slope, and the third implies that the stellar content of the 
``average'' early-type galaxy formed at high redshift.
A composite model---including variations in age and metallicity, as well as
a wavelength-independent effect such as homology breaking---is presented which
can fit all four observed properties.
This model implies that the most luminous early-type galaxies contain the oldest
and most metal-rich stars, while the lowest luminosity galaxies formed the bulk of
their stars as recently at $z_f \sim 1$.
\end{abstract}


\section{Introduction 		\label{FPintroduction}	}

The discovery that elliptical
galaxies follow small-scatter, bivariate
correlations among their properties (Dressler \arg{et al.}\ 1987; Djorgovski \&
Davis 1987)---which are now
referred to as the Fundamental Plane (FP)---provided a powerful new tool to 
study this galaxy population.
Not only do elliptical galaxies have only a small variation among their
properties for any given luminosity, but those mean properties also vary
smoothly and systematically from the faintest to the most luminous galaxies.
For a recent review, see Pahre \& Djorgovski (1997).

The bivariate FP correlations, and their monovariate projections (such as
the color-magnitude relation [CMR] or the Faber-Jackson relation), provide
the basic tool by which the evolution and origin of the elliptical galaxy 
population can be studied.
First, the evolution of the surface brightness ``intercept'' of the FP in
(see \S 4) is an indicator of the mean formation epoch 
for the stellar content of the ``average'' elliptical galaxy
(Barrientos \arg{et al.}\ 1996; Pahre \arg{et al.}\ 1996;
van Dokkum \& Franx 1996; Kelson \arg{et al.}\ 1997; Stanford \arg{et al.}\ 1998).
Second, the variation of the ``slope'' of the FP with wavelength 
(\S 3) and redshift (\S 4),
as well as its deviation from the virial expectation under the assumptions of homology
and constant $M/L$ (\S 2; Djorgovski \& Davis 1987; Faber \arg{et al.}\ 1987), 
are all direct evidence of \emph{systematic} variations in the intrinsic physical 
properties (age, metallicity, IMF, dark matter content, velocity anisotropy, homology
breaking) of ellipticals \emph{along} the galaxy sequence.
Third, the small scatter of the FP at any given luminosity implies that
all relevant, intrinsic physical properties, whether or not they are
varying along the galaxy sequence, possess only a small variation among all 
elliptical galaxies of a given luminosity.


The slope of the FP as a function of wavelength appears to be the key parameter 
for determining the physical origin of the elliptical galaxy sequence.
As was first pointed out more a decade ago (Djorgovski \& Davis 1987; Faber \arg{et al.}\ 1987),
the observed form of the FP deviates from the virial expectation (assuming homology and constant
$M/L$) of $R_{\rm eff} \propto \sigma_0^{2} \langle\Sigma\rangle_{\rm eff}^{-1}$ 
(under the assumptions of constant $M/L$ and homology among 
elliptical galaxies).
This effect was usually interpreted as due to a systematic breakdown in $M/L$ as a
function of galaxy luminosity,
which is, in turn, due to variations in
stellar or dark matter content along the elliptical galaxy sequence.
More recent studies have suggested that homology-breaking is occurring along the
galaxy sequence (Burkert 1993; Caon \arg{et al.}\ 1993;
Capelato \arg{et al.}\ 1995; Hjorth \& Madsen 1995; Busarello \arg{et al.}\ 1997;
Graham \& Colless 1997) which could produce part or all of the 
deviation of the observed FP from the virial expectation.
The answer may be a combination of the two (\S 7; Pahre \& Djorgovski 1997; 
Pahre \arg{et al.}\ 1998a,b).

Many past studies used simplistic models (e.g., elliptical galaxies form
\emph{only} a metallicity sequence) to explain a limited number of
observed properties [e.g., \emph{only} the $B$-band FP, or \emph{only}
the $(U-V)_0$ CMR].
As will be shown in \S 6, such simplistic models
are typically ruled out by considering other observed properties (enumerated in \S 5),
therefore requiring a more detailed, \emph{composite} model to explain all of 
the observations simultaneously.

\section{The Near-Infrared Fundamental Plane
	\label{near-IR-FP} }

If there exists a stellar populations component to the physical 
origin of the FP (as evidenced by its deviations from the virial
expectation), then the form of the FP should vary with wavelength
due to the stellar populations effects on the mean surface brightness
$\langle\mu\rangle_{\rm eff}$ which enters the FP.
Since $M/L_K$ is independent of metallicity (see the contribution 
by MARASTON in this volume), studies in the near-infrared can test 
the effects of metallicity on the slope of the FP.

The near-infrared FP described here is the result of a large survey of nearby 
galaxies in the $K$-band using IR imaging detectors (Pahre 1998b).
The near-infrared form of the FP is plotted in Figure~1,
and is represented by an equation of the form, 
\begin{equation}
R_{\rm eff} \propto \sigma_0^{1.53 \pm 0.08} \langle\mu\rangle_{\rm eff}^{-0.79 \pm 0.03}
\label{eq-pahre-kfp}
\end{equation}
(Pahre \arg{et al.}\ 1998a).
The slope of the FP therefore deviates from the virial expectation 
($R_{\rm eff} \propto \sigma_0^2$) even in the near-infrared.
Further implications of this scaling relation will be discussed below.

\begin{figure}
\centering \leavevmode \epsfxsize=0.5\columnwidth \epsfbox{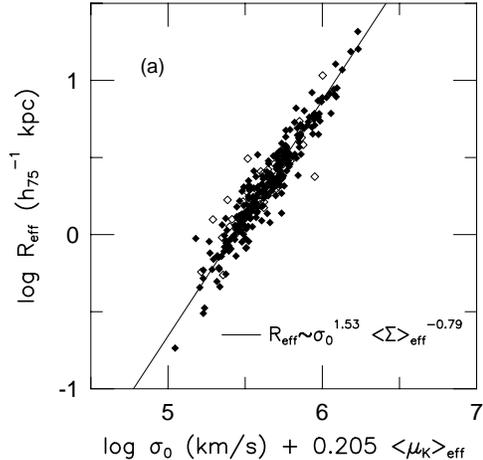}
\caption{The near-infrared form of the FP from Pahre \arg{et al.}\ (1998a).
	\label{fig-near-IR-FP} }
\end{figure}

\section{The Slope of the FP Changes With Wavelength
	\label{FP-slope-wavelength}	}

Early work on the FP argued that the deviation of the slope of the FP
from its virial expectation implied that the stellar $M/L$ varied weakly,
but systematically, as a function of galaxy mass or luminosity
(Djorgovski \& Davis 1987; Faber \arg{et al.}\ 1987).
Such stellar population effects---age, metallicity, or IMF---will have a
distinctive signature in the broadband colors of the galaxies, hence the
variations of the slope of the FP with wavelength ought to provide a
constraint on those stellar populations effects.

Variations among the optical bandpasses, however, were clearly small and not
substantially larger than the measurement uncertainties.
For the FP of the form $R_{\rm eff} \propto \sigma_0^a \langle\Sigma\rangle_{\rm eff}^b$,
Lucey \arg{et al.}\ (1991) found $(a,b)=(1.22 \pm 0.07,-0.83 \pm 0.06)$ in the $V$-band,
Djorgovski \& Davis (1987) found $(1.39 \pm 0.14,-0.90 \pm 0.09)$ in the $r_G$-band,
J\o rgensen \arg{et al.}\ (1996) found $(1.24 \pm 0.07,-0.82 \pm 0.02)$ in the $r$-band,
Scodeggio \arg{et al.}\ (1997) found $(1.25 \pm 0.02,-0.80 \pm 0.03)$ in the $I_{\rm C}$-band,
and Pahre \arg{et al.}\ (1998a) found $(1.53 \pm 0.08,-0.79 \pm 0.03)$ in the $K$-band.
The wavelength dependence of the slope of the FP is small and not accurately constrained from
these measurements.
Clearly, an improved method to measure the change in the slope of the FP with wavelength 
is desired to quantify accurately the stellar populations contributions to the slope of the FP
at all wavelengths.


The change in slope of the FP from the $V$ to the $K$-band 
can be calculated by setting the SB coefficient to $0.32$
\begin{eqnarray}
r_{\rm eff,V} & = & a_V \log \sigma_0 + 0.32 \langle\mu_V\rangle_{\rm eff} + {\rm~constant} \\
r_{\rm eff,K} & = & a_K \log \sigma_0 + 0.32 \langle\mu_K\rangle_{\rm eff} + {\rm~constant}
\end{eqnarray}
and then taking the difference
\begin{equation}
(\log r_{\rm eff,K} - 0.32 \langle\mu_K\rangle_{\rm eff}) 
	- (\log r_{\rm eff,V} - 0.32 \langle\mu_V\rangle_{\rm eff}) 
	= \Delta a \log \sigma_0 + {\rm constant} ,
\label{eq-pahre-deltafp}
\end{equation}
where $\Delta a = a_K - a_V$ is the change in the slope of the FP.
Note that both sides of Equation~4 are distance-independent quantities; 
the analysis that follows therefore works both in clusters and the general field.
If a particular galaxy sample suffers, say, from the resolved depth
of a cluster, this approach will remove such distance-dependent effects
on $r_{\rm eff}$ from the analysis.
Furthermore, this approach requires that a given galaxy have observations
in both observed bandpasses, so there are no longer differences in the
definition of two given galaxy samples---only those galaxies in common between
the two datasets are used.

\begin{figure}
\centering \leavevmode \epsfxsize=.35\columnwidth \epsfbox{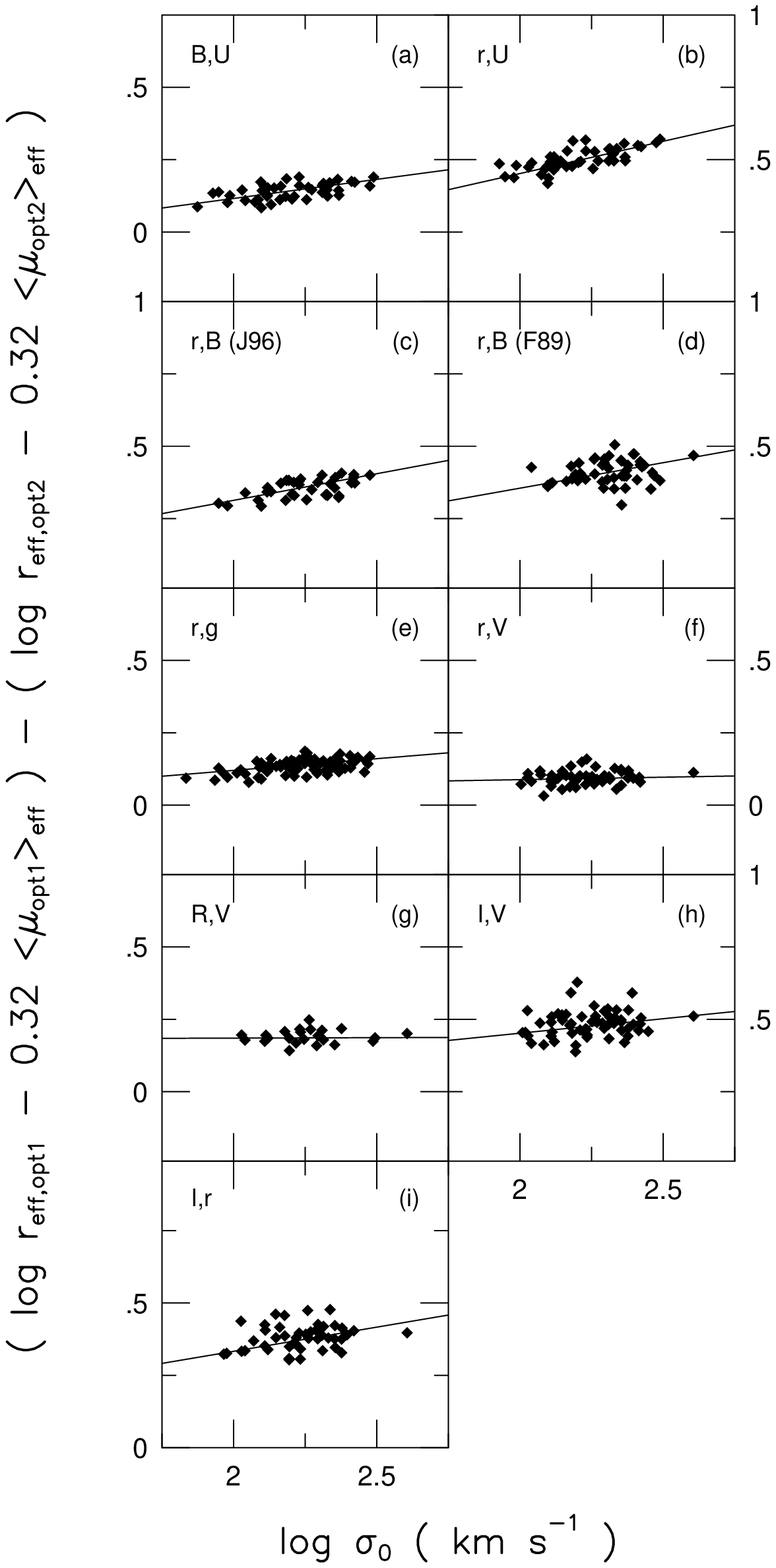} \hfil
\epsfxsize=.35\columnwidth \epsfbox{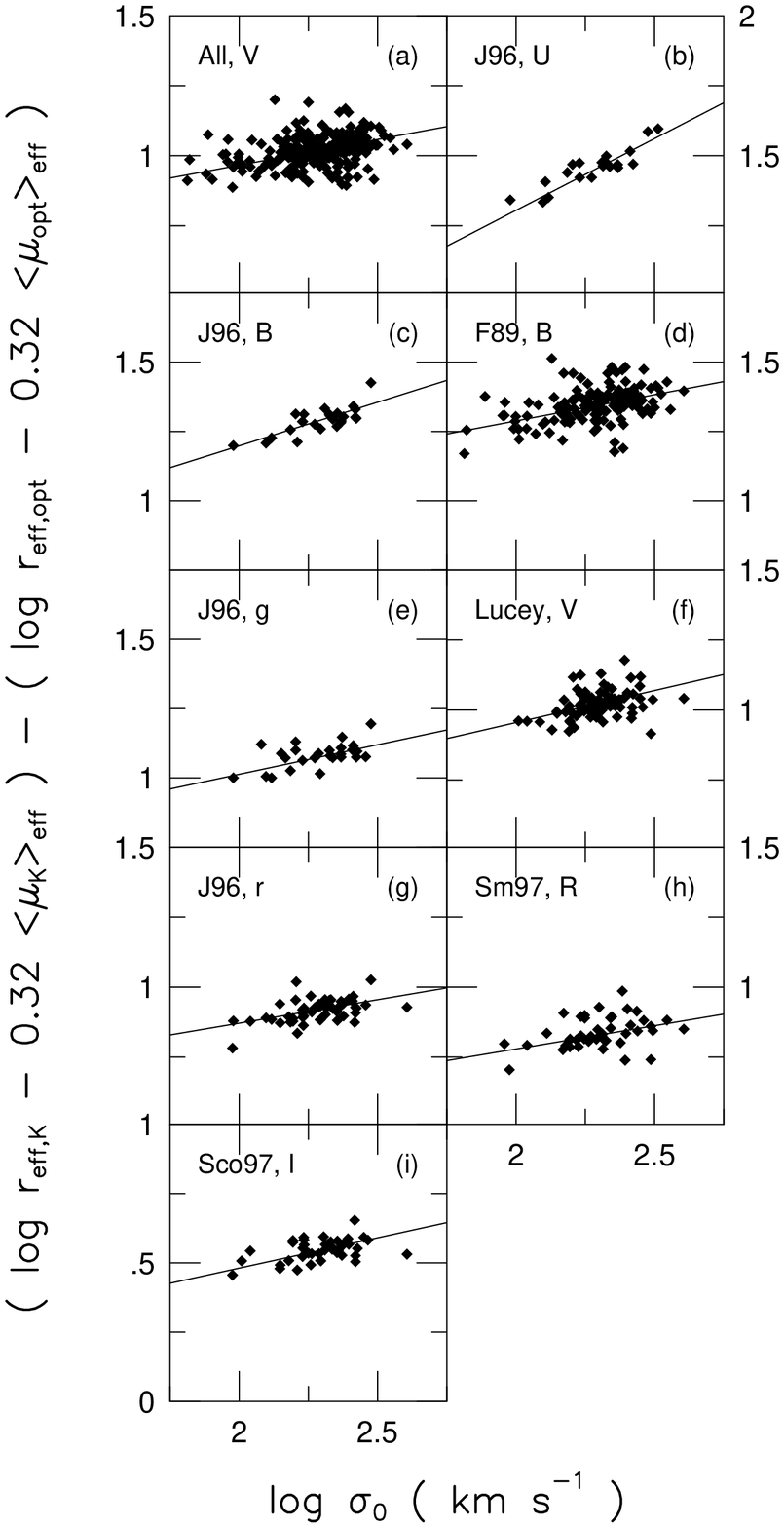}
\caption{The change in the slope of the FP between pairs of bandpasses
[left] at optical wavelengths, and [right] optical and near-infrared wavelengths.
If the slope of the FP were independent of wavelength, then the data in each panel would
lie along a horizontal line; 
virtually every comparison, however, shows a statistically significant positive regression,
indicative that the slope of the FP is increasing with wavelength.
(From Pahre \arg{et al.}\ 1998b.)
	\label{fig-FP-wavelength} }
\end{figure}

The generalized form of Equation~4 for various pairs
of bandpasses are plotted in Figure~2 (from Pahre \arg{et al.}\ 1998b),
which demonstrates that the slope of the FP increases with wavelength for
virtually every data comparison in the optical and infrared.
The conclusion that the slope of the FP increases with wavelength is a strong 
indication that stellar populations variations contribute to the slope and
hence help to define the early-type galaxy sequence.

\section{The Evolution of the FP With Redshift
	\label{FP-evolve-redshift} }

Since the FP represents the elliptical galaxy scaling relation with the
smallest possible scatter, it is, by construction, the optimal tool to
study the changes of the elliptical galaxy population with redshift.
The color-magnitude relation has gained great popularity to study elliptical
galaxy evolution, but the inclusion of the central velocity dispersion
term of the FP allows for the estimation of dynamical mass.
It is for this reason that the evolution of the FP is often referred to
as the evolution of the mass-to-light ratio---although it is really only
the \emph{evolution of galaxy luminosity at fixed mass.}

\begin{figure}
\centering \leavevmode \epsfxsize=0.9\columnwidth \epsfbox{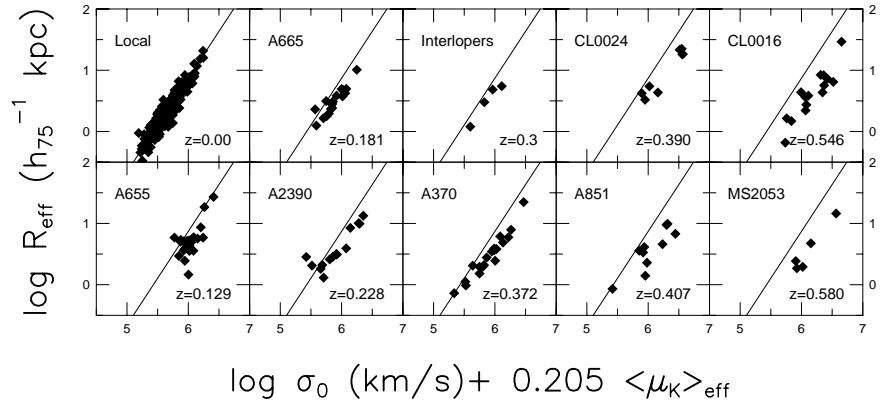}
\caption{The FP for $0 < z < 0.6$, with each individual panel representing
	a different cluster.  The local galaxies (Pahre \arg{et al.} 1998a) are 
	in the upper left panel.  (From Pahre 1998b.)
	\label{fig-highz-FP-Kband} }
\end{figure}

The field was pioneered by the velocity dispersion measurements at $z=0.18$
of Franx (1993) and at $z=0.39$ by van Dokkum \& Franx (1996).
Those observations at the MMT used very long exposure times to obtain the
necessary S/N to measure the velocity dispersions.
The advent of the Keck 10~m Telescope with its high-throughput spectrograph
(Oke \arg{et al.} 1995) allows for higher S/N spectra to be obtained in
far less observing time, which means that real \emph{surveys} of the properties 
of early-type galaxies at $0 < z < 1$ are possible.
For example, the \emph{combined} studies of van Dokkum \& Franx (1996), 
Kelson \arg{et al.}\ (1997), Ziegler \& Bender (1997), and 
van Dokkum \arg{et al.} (1998a) document only 36 galaxies at $z > 0.1$ that are
suitable for studies of the FP.
Our new survey with the Keck Telescope adds more than 100 galaxies 
at $0.1 < z < 0.6$ to create a large, homogeneous dataset 
for the evolutionary studies.
The galaxies for each cluster were selected using two-color (three bandpass) 
and morhpology (concentration index) criteria as described in Pahre (1998a).

The evolution of the SB intercept of the FP should probe the mean luminosity 
evolution of the ``average'' early-type galaxy, while the evolution of the 
slope of the FP should probe the relative evolution of the bright galaxies
when compared to the faint galaxies.
Only the evolution of the intercept has been discussed previously; the large
galaxy sample described here allow the slope to be constrained for the first
time at these redshifts.
The $K$-band FP for eight clusters at $0.1 < z < 0.6$ are presented in 
Figure~3.

Several effects are evident.
First, the galaxies at $z > 0.1$ all lie to the right of the relation for
the nearby galaxies, which is due to the cosmological SB dimming effect.
A small amount of luminosity evolution is also present, which moves the data
points a small amount back to the left in each panel.
This is clear after the SB dimming effect is removed, and is plotted in
Figure~4 [left].
Finally, the slope of the FP relation appears to become slightly flatter
with redshift.
This effect is quantified in Figure~4 [right],
where the slope is measured to become smaller with redshift.

\begin{figure}
\centering \leavevmode \epsfxsize=.45\columnwidth \epsfbox{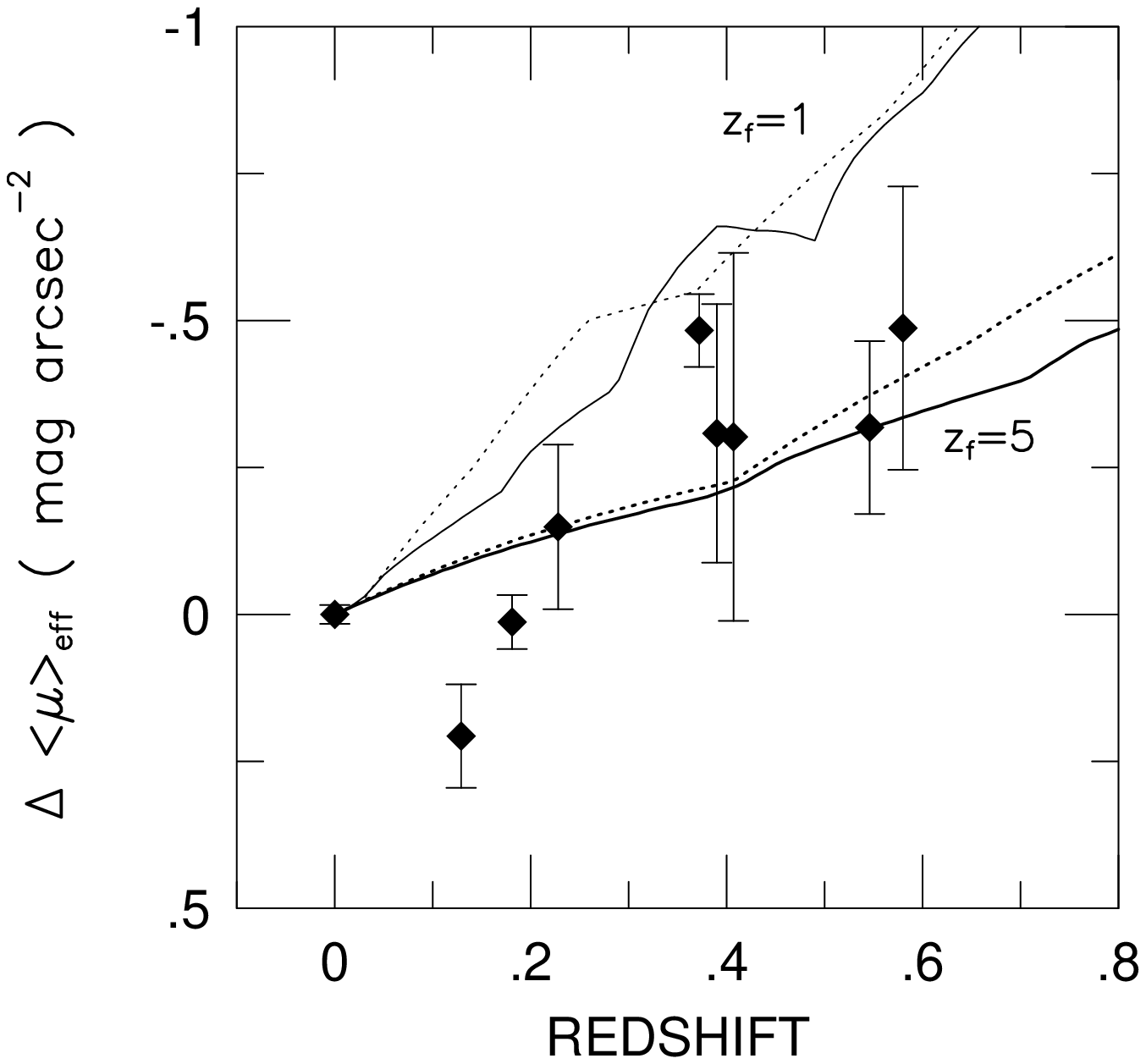} \hfil
\epsfxsize=.43\columnwidth \epsfbox{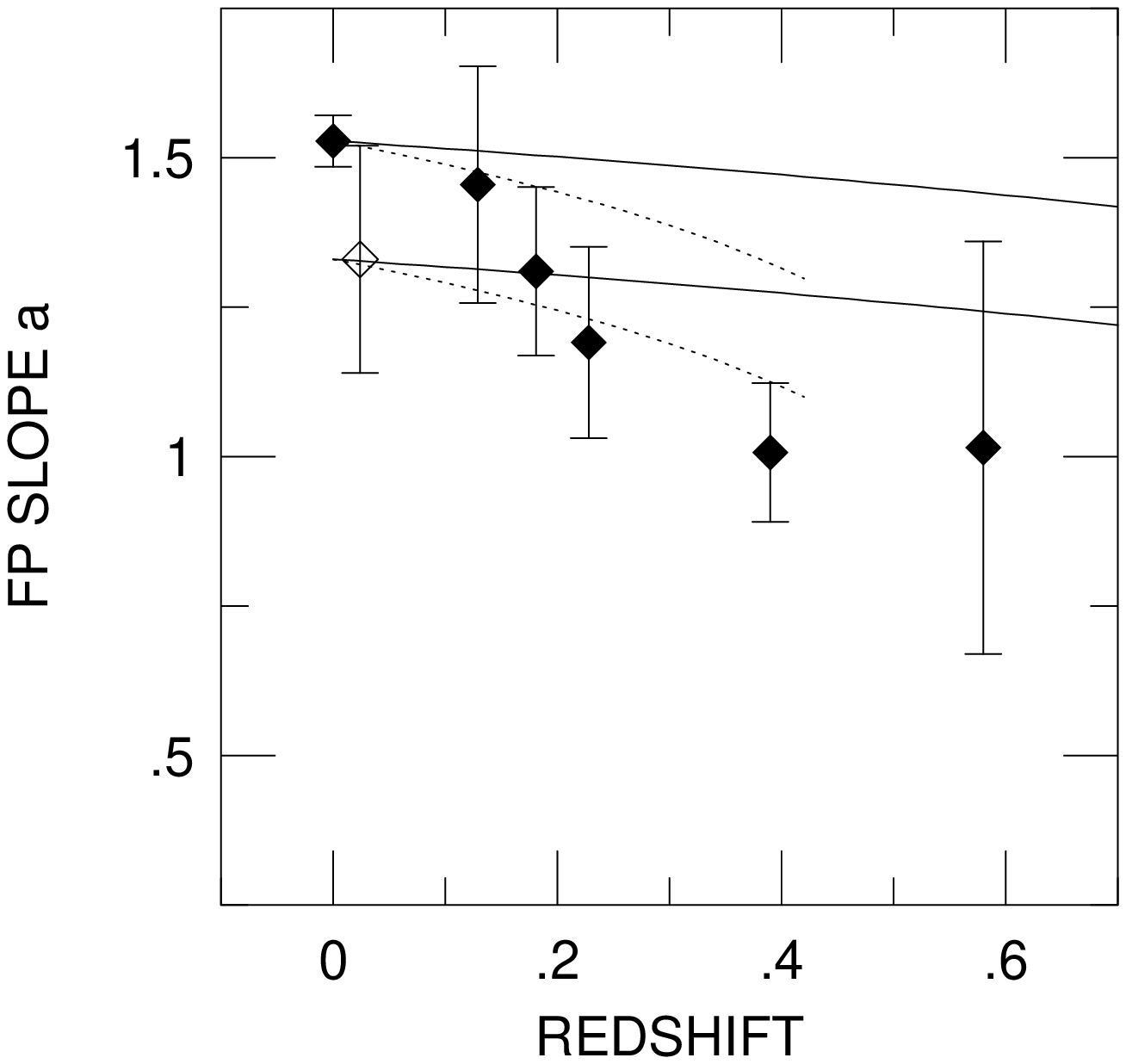}
\caption{The evolution of the surface brightness intercept [left] and the
	slope [right] of the $K$-band FP with redshift.  (From Pahre 1998a.)
	\label{fig-FP-evolution} }
\end{figure}

It is apparent from the luminosity evolution plot in Figure~4
that the ``average'' early-type galaxy is evolving passively as though its stars
were formed at high redshift $z_f \geq 3$.
The evolution of the slope of the FP with redshift, however, argues that there
is a variation in formation redshift along the early-type galaxy sequence:
the brightest early-type galaxies formed their stars first at $z_f \geq 5$, 
while the  faintest early-type galaxies formed their stars later---as late 
as $z_f \sim 1$.
The model lines overplotted in Figure~4 [RHS] are the \emph{predictions}
based on the comprehensive models based on the nearby galaxies 
(see \S 7); two different normalizations are given 
(either to all the local galaxies, or to the Coma cluster alone; for details 
see Pahre 1998a), which provides the information as the the age spreads allowed 
by the evolution of the slope of the FP.

Several other lines of evidence may also show this effect.
van Dokkum \arg{et al.} (1998b) showed that the color-magnitude relation for
S0 galaxies in a cluster at $z \sim 0.3$ are typically bluer than the E galaxies.
Kuntschner \& Davies (1998) used line indices to show that the S0 galaxies 
in the Fornax cluster have a range of ages, while the elliptical galaxies 
show very little variation in age.
Both of these results could be highlighting the same population that is 
causing the evolution of the FP slope with redshift seen here, since S0 galaxies
are typically less luminous than E galaxies.
Finally, the color-magnitude relation observations at $0.3 < z < 0.9$ in
Stanford \arg{et al.}\ (1998) used large apertures to measure the galaxy
colors; if there is a correlation between the size of the populations 
gradient and the galaxy luminosity, as is suggested by Gonz\'alez \& Gorgas
(1995; see also the contribution by GONZ\'ALEZ to this volume), then this 
effect would cause the slope of the \emph{corrected} color-magnitude 
relation to evolve with redshift, as appears to have been found in an 
independent analysis (see contribution by KODAMA in this volume).\footnote{Note
that Kodama has reached a different conclusion from the present work for the
origin of evolution of the CMR slope in the data of Stanford \arg{et al.}, 
although it is not yet clear to what extent composite models of age and metallicity 
are allowed by his analysis. }

\section{Constraints on the Origin of the Early-Type Galaxy Sequence
	\label{egalaxy-sequence} }

In the course of this meeting, there were many arguments presented as to
the physical origins of the elliptical galaxy sequence.
There are a number of observed properties of elliptical galaxies which should be 
explained by any proposed model for the origin of the galaxy sequence:
\newcounter{properties}
\begin{list}{\arabic{properties}.}{\topsep=0in \itemsep=-0.05in \usecounter{properties}}
\item the slope of the FP increases with wavelength;
\item the intercept of the FP and CMR evolve with redshift in an apparently passive manner;
\item the slope of the $K$-band FP deviates from the virial expectation (assuming constant
	$M/L$ and homology);
\item the slope of the FP flattens with redshift;
\item the slope of the CMR increases with redshift; 
\item the residuals of the elliptical galaxy correlations do not show clear
	trends with age or metallicity; and
\item the velocity distributions of elliptical galaxies appear to deviate from
	a homologous scaling family.
\end{list}
To these can be added a general property of galaxies in clusters:
\begin{list}{\arabic{properties}.}{\topsep=0in \itemsep=-0.05in \usecounter{properties}}
\setcounter{properties}{7}
\item the fraction of blue galaxies in clusters 
	increases with redshift (Butcher-Oemler effect).
\end{list}



\section{Simple Models for the Origin of the FP
	\label{FP-simple-models}	}

\noindent \textbf{Metallicity Sequence.}  
Most researchers working in this field would probably wager that elliptical galaxies
form a metallicity sequence, with the most luminous galaxies also being the most metal-rich.
This would naturally explain the change in the slope of the FP with wavelength 
in \S 3.
Even though a metallicity sequence seems to match the evolution of the color-magnitude
relation (see the contribution by KODAMA in this volume), it cannot explain the $K$-band FP:
such a metallicity sequence model \emph{must also} invoke either homology breaking
or systematic variations in dark matter content or its distribution.\footnote{(See 
the contribution by LOEWENSTEIN in this volume, which includes a discussion of the 
scale-length changes for the dark matter distribution as a function of galaxy luminosity.)}
The metallicity sequence also fails to explain the evolution of the slope of the FP with 
redshift, and provides no explanation for the Butcher-Oemler galaxies at intermediate
redshifts or their present-day counterparts.

\noindent \textbf{Systematic Homology Breaking.}  
It is currently popular to explore systematic homology breaking as the origin of the FP.
This effect, however, is \emph{independent of wavelength}, which contradicts the variation
of the FP slope with wavelength shown in \S 3.
Systematic homology-breaking along the elliptical galaxy sequence cannot \emph{by itself}
explain the origin of the FP at all wavelengths without also invoking systematic variations
in stellar content along the FP.

\noindent \textbf{Conspiracy Theories.}  
One possibility is that variations in metallicity could offset those in age in order
to keep the FP thin.
A model of this kind was proposed by Worthey \arg{et al.}\ (1995), who predicted
that $M/L_K \approx {\rm constant}$.
The deviation of the $K$-band FP from the virial expectation implies 
$M/L \propto M^{0.15 \pm 0.01}$, which contradicts this model.

\begin{figure}
\centering \leavevmode \epsfxsize=0.5\columnwidth \epsfbox{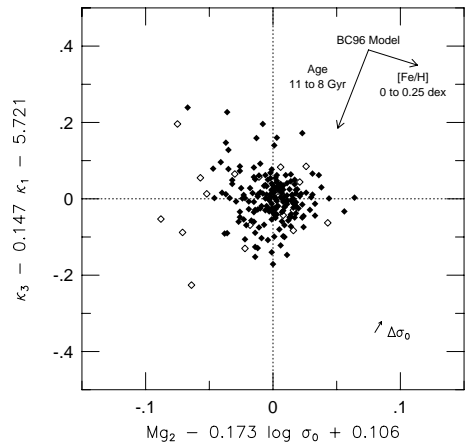}
\caption{Residuals from the $\kappa$-space form of the $K$-band FP [vertical axis] 
	plotted against the residuals from the Mg$_2$--$\sigma_0$ relation [horizontal axis].
	The residuals show no correlation in the directions of either age or metallicity,
	suggesting that there is no single origin for the scatter of the scaling relations.
	\label{fig-deltakappa-deltamg2sigma} }
\end{figure}

\noindent \textbf{Scatter of the FP.}  
One explanation for the scatter of the FP is that it is due to small
metallicity or age variations among elliptical galaxies at any given luminosity.
Since $M/L_K$ is a function of age but not metallicity, and Mg$_2$ is primarily
a function of metallicity but not age (Mould 1978), the residuals of the $K$-band
FP and the Mg$_2$--$\sigma_0$ relation should provide a clean separation of age
and metallicity effects on the scatter of the relations.
These are plotted in Figure~5, and show no clear trend with either age or metallicity.
The origin of the scatter of the FP is not so simply described:  it probably
has an origin in both age and metallicity, and possibly also in another effect
(like dissipationless merging).

\section{A Comprehensive Model for the Early-Type Galaxy Sequence
	\label{FP-comprehensive-model}	}

The deviation of the $K$-band FP from the virial expectation (Figure~1)
implies that metallicity variations alone are insufficient to explain the origins of the FP;
age and/or a wavelength-independent effect (homology breaking or systematic variations
in dark matter content) are also required.
The variation of the slope of the FP with wavelength (Figure~2) 
implies that homology breaking or dark matter variations alone are insufficient to explain 
the origins of the FP---stellar populations are also required.
The lack of a correlation among the residuals of the $K$-band FP and the 
Mg$_2$--$\sigma_0$ relation (Figure~5) implies
that the scatter of the FP cannot be attributed to age or metallicity effects alone.
A complete understanding of the physical origins of the FP will therefore require
a detailed model: it must also simultaneously account for all of the elliptical galaxy
correlations at all redshifts.

Such a model has been constructed by Pahre \arg{et al.}\ (1998b), which provides
four model parameters:  
(1) variation in age from one end of the elliptical galaxy sequence to the other; 
(2) variation in metallicity along the entire galaxy sequence;
(3) a mean value for stellar populations gradients;
and (4) a wavelength-independent effect such as non-homology.
Simple stellar populations models (Bruzual \& Charlot 1996, in preparation, as appeared
in Leitherer \arg{et al.}\ 1996; Vazdekis \arg{et al.}\ 1996)
are used to convert between observed and physical quantities.
This is a decidedly empirical model that is meant to \emph{describe} the global
properties of elliptical galaxies in a complete manner, but not explain the origins
of these effects in the galaxy formation process.

This model has a range of observational measurements that can be used as equations
of constraint:
(1) the slope of the $K$-band FP; 
(2) the variations in the slope of the FP with wavelength;
(3) the slope of the Mg$_2$--$\sigma_0$ relation;
(4) a mean value for color gradients in ellipticals;
and (5) a measurement of the aperture effect on velocity dispersion.
The effects of the stellar populations gradients on the observed
parameters is fully accounted for in the model.

This analysis demonstrates that both age and metallicity are varying along
the elliptical galaxy sequence, with the relative contributions of the two
depending on the simple stellar populations model used---up to a factor of
two in age and a factor of three in metallicity.
The variations are in the sense that the most luminous galaxies are 
the oldest and the most metal-rich; the latter effect appears to contradict
the conclusions of Trager (1997; see, however, the discussion in Pahre 1998a).

The result that age contributes to the slope of the FP suggests that the
slope of the FP should evolve with redshift, since the lowest luminosity
galaxies are the youngest and hence will evolve fastest with redshift.
This prediction, based solely on the properties of the nearby galaxies,
is plotted in Figure~4 against the observed evolution 
of the slope of the FP with redshift.


\acknowledgments

M.~A.~P. is supported by Hubble Fellowship grant HF-01099.01-97A from STScI 
(which is operated by AURA under NASA contract NAS5-26555).
Additional partial support was provided for this work by AR grant GO-6381
from STScI and the Bressler Foundation.
Many thanks to the conference organizers for putting together a fun and
lively meeting.


\end{document}